\documentclass[12pt]{article}%
\usepackage{graphicx}
\usepackage{amsmath}
\usepackage{amsfonts}
\usepackage{amssymb}%
\begin{document}

\title{ Theoretical analysis of resonance states in $^{4}H$, $^{4}He$ and $^{4}Li$
above three-cluster threshold}
\author{V. Vasilevsky$^{\dagger\ddag},$ F. Arickx$^{\ddag}$, J. Broeckhove$^{\ddag}$,
and V.N.Romanov$^{\dagger}$\\$^{\dag}$ Bogolyubov Institute for Theoretical Physics,\\Kiev, Ukraine and\\$^{\ddag}$ Departement Wiskunde en Informatica, Universiteit Antwerpen (RUCA),\\Antwerpen, Belgium}
\date{}
\maketitle

\begin{abstract}
The resonance states of $^{4}H$, $^{4}He$ and $^{4}Li$, embedded in the
three-cluster $d+N+N$ continuum, are investigated within a three-cluster
model. The model treats the Pauli principle exactly and incorporates the
Faddeev components for proper description of the boundary conditions for the
two- and three-body continua. The hyperspherical harmonics are used to
distinguish and numerate channels of the three-cluster continuum. It is shown
that the effective barrier, created by three-cluster configuration $d+N+N$, is
strong enough to accommodate two resonance states.

\end{abstract}

\input epsf

\begin{flushleft}
UDC 539.142
\end{flushleft}

\section{{\protect\large Introduction}}

In this paper we study the nature of resonance states in $^{4}H$, $^{4}He$ and
$^{4}Li$. All these nuclei have a rich structure of resonance states
\cite{Tilley1992A4}. There are 4 well-determined resonances in $^{4}H$ and
$^{4}Li$, and up to 15 resonance states were detected in $^{4}He$. Most of
these resonances have a width that is much larger than the resonance energy
when measured from the lowest threshold. Although these resonances have been
experimentally confirmed, they can hardly  be observed in current theoretical
model descriptions of these systems through standard elastic and inelastic
scattering quantities such as $S$-matrix elements, differential or total cross
sections, and so on.

For many years, the four-nucleon system was studied by different microscopic
and semi-microscopic methods. Different forms of the Schr\"{o}dinger equation
(differential (\cite{2001VI05, 2001VI09}), integral (\cite{2000FI02,
Carbonell2001, Carbonell200}), integro-differential {\cite{2002KA42,2001PF02,
1991PhRvC..43..371K, 2002PThPh.107..833K}, matrix or
algebraic(\cite{kn:vv4he_3chE,kn:VVS+19904HeE,1997HO01}), ...) have been used
to study these nuclei. Special attention was paid to $^{4}He$, the only
nuclear 4-particle system featuring a bound state. The theoretical study of
the \textquotedblleft ground state\textquotedblright\ of $^{4}H$ and $^{4}Li$,
and of the excited states of all three nuclei were investigated mainly through
resonance state analysis. Of all resonances, the first excited $0^{+} $ state
has received most attention. In none of the descriptions the role of the
three-cluster channels was properly considered, and so resonance states
induced by this channel could not be theoretically discovered and analyzed. }

Our objective is to determine the type and nature of resonance states in
$^{4}H$, $^{4}He$ and $^{4}Li$ that are reproduced within a three-cluster
microscopic model. For all three nuclei we will consider one single
three-cluster configuration $d+N+N$. This is certainly the most dominant
three-cluster channel, as it has the lowest energy threshold. Moreover one can
easily construct all binary channels for these nuclei within such description.
In $^{4}He$ for example this configuration allows to study resonances created
by the two-cluster channels $p+^{3}H$, $n+^{3}He$ and $d+d $, as well as by
the three-cluster channel\ $d+p+n$. The latter should be very important,
because 7 resonance states were detected above the $d+p+n$ threshold. It is
interesting to note in the same context that all four resonance states of
$^{4}H$ lie below the three-cluster $d+n+n$ threshold, whereas in $^{4}Li$ two
resonances are found above the $d+p+p$ threshold.

We propose a modification of the method formulated in \cite{kn:vasil97e,
A6Proc, kn:ITP+RUCA1}, and used in \cite{kn:ITP+RUCA2, kn:ITP+RUCA3} to study
resonances embedded in the three-cluster continuum, and reactions with three
cluster exit channels. The method was shown to provide a suitable instrument
for investigating Borromian nuclei (for instance $^{6}He$) and nuclei with
prominent three-cluster features (like $^{6}Be$). We wish to extend the method
proposed in \cite{kn:vasil97e, kn:ITP+RUCA1} to handle systems in which binary
channels play a prominent role, by including the correct boundary conditions
for both binary and ternary channels. The results obtained in
\cite{kn:vasil97e, VNCh2001E} and in \cite{kn:ITP+RUCA2, kn:ITP+RUCA3} allow
us to restrict the model space to the most relevant subspace.

To reach this objective we have to:

\begin{itemize}
\item specify the microscopic modelling of the three-cluster configuration,
and the approximations to be used in calculations,

\item construct a set of dynamic equations that take into account the\ proper
boundary conditions for both binary and ternary channels,

\item implement reliable numerical methods to calculate continuous spectrum
wave functions and $S$-matrix elements.
\end{itemize}

\section{{\protect\large Model and Methodology}}

We propose the following trial wave function for the 4-particle systems%
\begin{align}
\Psi &  = \Psi_{1}+\Psi_{2}+\Psi_{3}\label{eq:a001}\\
&  = \widehat{\mathcal{A}}\left\{  \Phi_{1}\left(  A_{1}\right)  \Phi
_{2}\left(  A_{2}\right)  \Phi_{3}\left(  A_{3}\right)  \left[  f_{1}\left(
\mathbf{x}_{1},\mathbf{q}_{1}\right)  +f_{2}\left(  \mathbf{x}_{2}%
,\mathbf{q}_{2}\right)  +f_{3}\left(  \mathbf{x}_{3},\mathbf{q}_{3}\right)
\right]  \right\}  , ~\nonumber
\end{align}
where $\Phi_{\alpha}\left(  A_{\alpha}\right)  $ is the antisymmetric and
translationally invariant internal wave function of the $A_{\alpha}$\ nucleon
system, and $\mathbf{x}_{\alpha},\mathbf{q}_{\alpha}$ are the familiar Jacobi
coordinates denoting ($\mathbf{x}_{\alpha}$) the relative position of two of
the clusters, and ($\mathbf{q}_{\alpha}$) the relative position of the third
cluster with respect to the former two-cluster subsystem%
\begin{align}
\mathbf{x}_{\alpha}  &  =\sqrt{\frac{A_{\beta}A_{\gamma}}{A_{\beta}+A_{\gamma
}}}\left[  \frac{\sum_{j\in A_{\beta}}\mathbf{r}_{j}}{A_{\beta}}-\frac
{\sum_{k\in A_{\gamma}}\mathbf{r}_{k}}{A_{\gamma}}\right]  ,\nonumber\\
\mathbf{q}_{\alpha}  &  =\sqrt{\frac{A_{\alpha}\left(  A_{\beta}+A_{\gamma
}\right)  }{A_{\alpha}+A_{\beta}+A_{\gamma}}}\left[  \frac{\sum_{i\in
A_{\alpha}}\mathbf{r}_{i}}{A_{\alpha}}-\frac{\sum_{j\in A_{\beta}}%
\mathbf{r}_{j}+\sum_{k\in A_{\gamma}}\mathbf{r}_{k}}{A_{\beta}+A_{\gamma}%
}\right]  ,\nonumber
\end{align}
with $\left(  \alpha,\beta,\gamma\right)  $ a cyclic permutations of
$(1,2,3)$. The three components of the wave functions $\left\{  \Psi_{1}%
,\Psi_{2},\Psi_{3}\right\}  $ (more precisely $\left\{  f_{1},f_{2}%
,f_{3}\right\}  $) have to be determined by solving the many-particle
Schr\"{o}dinger equation. Specific symmetries of the system can reduce the
number of components: if the three-cluster configuration contains two
identical clusters, only two distinguishable components $\left\{  f_{1}%
,f_{2}\right\}  $ will occur; this is the case for $^{4}H$ and $^{4}Li$. If
all three clusters are identical (impossible for the 4-particle system
though), only one independent component $\left\{  f_{1}\right\}  $ would occur.

We shall use a matrix or algebraic form of the Schr\"{o}dinger equation. To
this aim we expand the wave function $f_{\alpha}\left(  \mathbf{x}_{\alpha
},\mathbf{q}_{\alpha}\right)  $ in an oscillator basis (referred to as a
BiOscillator\ (BO) basis):%
\begin{equation}
f_{\alpha}\left(  \mathbf{x}_{\alpha},\mathbf{q}_{\alpha}\right)  =\sum
_{n_{y},l,n_{x},\lambda}C_{n_{y},l,n_{x},\lambda}^{\left(  \alpha\right)
}\left\vert n_{y},l,n_{x},\lambda;LM\right\rangle _{\alpha}, \label{eq:a003}%
\end{equation}
where%
\begin{equation}
\left\vert n_{y},l,n_{x},\lambda;LM\right\rangle _{\alpha}=\Phi_{n_{y}%
,l}\left(  q_{\alpha}\right)  \Phi_{n_{x},\lambda}\left(  x_{\alpha}\right)
\left\{  Y_{l}\left(  \widehat{\mathbf{q}}_{\alpha}\right)  \cdot Y_{\lambda
}\left(  \widehat{\mathbf{x}}_{\alpha}\right)  \right\}  _{LM} \label{eq:a004}%
\end{equation}
and $\Phi_{n,l}\left(  q\right)  $ is the familiar (radial) oscillator
function:%
\begin{equation}
\Phi_{n,l}\left(  q\right)  =\left(  -1\right)  ^{n}\sqrt{\frac{2\Gamma\left(
n+1\right)  }{\Gamma\left(  n+l+3/2\right)  }}\frac{1}{b^{3/2}}\left(
\frac{q}{b}\right)  ^{l}\exp\left\{  -\frac{1}{2}\left(  \frac{q}{b}\right)
^{2}\right\}  L_{n}^{l+1/2}\left(  \left(  \frac{q}{b}\right)  ^{2}\right)  .
\label{eq:a005}%
\end{equation}
The total angular momentum $L$ of the three $s$-clusters is a vector sum of
two partial angular momenta $l_{\alpha}$ and $\lambda_{\alpha}$ associated
with the respective Jacobi coordinates $\mathbf{x}_{\alpha}$ and
$\mathbf{q}_{\alpha}$.

As each cluster function $\Phi_{\alpha}\left(  A_{\alpha}\right)  $ is
antisymmetric by construction, the antisymmetrization operator in
(\ref{eq:a001}) only involves the permutation of nucleons between clusters,
and it can be represented as%
\begin{equation}
\widehat{\mathcal{A}}=1+\widehat{\mathcal{A}}_{12}+\widehat{\mathcal{A}}%
_{23}+\widehat{\mathcal{A}}_{31}+\widehat{\mathcal{A}}_{123}, \label{eq:a006}%
\end{equation}
where $\widehat{\mathcal{A}}_{\alpha\beta}$ exchanges nucleons from clusters
$\alpha$\ and $\beta$, and $\widehat{\mathcal{A}}_{123}$ permutes particles
from all three clusters. In some respects this antisymmetrization operator is
similar to a short range interaction. Indeed, the antisymmetrization is
noticeable only when the distance between clusters is small. At larger
distances both the potential energy and the antisymmetrization effects become
negligibly small. The operator $\widehat{\mathcal{A}}_{123}$ is important only
when the distance between all three clusters is less than a certain restricted
value. If one of the clusters (say, cluster $\alpha$) is far away from the two
other clusters $\beta$ and $\gamma$, and the latter are close to one other,
$\widehat{\mathcal{A}}_{\beta\gamma}$ will have a pronounced contribution, as
well as the two-cluster interaction $\widehat{V}_{\beta\gamma}$ derived from
the $NN$-potential.

Each set of Jacobi coordinate $\mathbf{x}_{\alpha}$ and $\mathbf{q}_{\alpha}$,
and each set of oscillator functions (\ref{eq:a004})
\begin{equation}
\left\{  \left\vert n_{y},l_{\alpha},n_{x},\lambda_{\alpha};LM\right\rangle
_{\alpha}\right\} \nonumber
\end{equation}
for $\alpha=$ 1, 2 or 3 cover the whole configuration space (i.e. account for
all possible relative positions of three clusters in space). We will limit
ourselves to the subspace $\left\{  \left\vert n_{y},l_{\alpha}=L,n_{x}%
,\lambda_{\alpha}=0;LM \right\rangle \right\}  $ of the total Hilbert space.
Two arguments for such a choice can be given (in particular for the 4-particle system):

\begin{enumerate}
\item We deal with $s$-shell clusters, and the two-cluster compound subsystems
also are $s$-shell nuclei; the latter ($d$, $^{3}H$ and $^{3}He)$ have a
ground state containing a dominant $S$-wave contribution.

\item It was shown in \cite{A6Proc, kn:vasil97e, VNCh2001E} that this subspace
dominates in the full Hilbert space. For instance, the ground state energy of
$^{6}He$ and $^{8}He$ obtained within this subspace differs less than 1\% from
the energy obtained in the full Hilbert space. It was also shown that this
subspace dominates in the wave function of the $2^{+}$ state of $^{6}He$,
appearing as a resonance embedded in the three-cluster continuum $\alpha+n+n$,
as well.
\end{enumerate}

To include the proper boundary conditions, we will split the oscillator space
$\left\{  C_{n_{y},l,n_{x},\lambda}^{\left(  \alpha\right)  }\right\}  $
($\alpha=1,2,3$) into the internal and asymptotic parts. The former consists
of the basis functions of the lowest $N_{i}$ oscillator shells (i.e. all
functions with $N=n_{y}+n_{x}=0,1,2,\ldots N_{i}$; it involves $\left(
N_{i}+1\right)  \left(  N_{i}+2\right)  /2$ basis functions for each value of
$\alpha$). With this size of the internal region in oscillator space one can
evaluate the maximal size of the three-cluster system in coordinate space by
using the correspondence between oscillator and coordinate representations
(see details in \cite{kn:Fil_Okhr, kn:Fil81})%
\begin{equation}
R_{\max}\simeq b\sqrt{4N_{i}+6}.
\end{equation}

If, for instance, the oscillator length $b=1.5$ fm and $N_{i}=15$, then
$R_{\max}\simeq12$ fm. The maximal distance between any pair of clusters will
be of the same magnitude. $R_{\max}$\ also corresponds to the minimal size of
the three-cluster system in the asymptotic region. So in the internal region
all three clusters are close to each other, which means that all
antisymmetrization components are important, as well as all interactions
between clusters.

In the asymptotic region we distinguish two different regimes. In the first
regime the distance between two clusters is small, while the third one is far
apart. In the second regime all three clusters are well separated. If a
selected pair of clusters (say, $\beta$\ and $\gamma$\ clusters) has (a) bound
state(s), then the first regime is responsible for scattering of the third
cluster on the compound $\beta+\gamma$ subsystem. This process can be
described by two-body asymptotics. The second regime is associated with the
full disintegration of the three-cluster system, with three independent
(non-interacting) clusters. These two regimes have to be treated differently.
This means that two different forms of the wave function have to be used to
properly describe these processes. It will require some reconstruction of the
basis functions in order to suit both two- and three-body physical processes
in the exit channels.

In the first regime of the asymptotics we can neglect all antisymmetrization
components but $\widehat{\mathcal{A}}_{\beta\gamma}$. As for the potential
energy, the most important contribution is generated by the two-cluster
potential $\widehat{V}_{\beta\gamma}$. The other components $\widehat
{V}_{\alpha\beta}$ and $\widehat{V}_{\alpha\gamma}$, originating from the
short-range $NN$-forces, are negligibly small, and only long-range Coulomb
forces are of importance. The wave function in coordinate and oscillator
representation can then be factorized as%
\begin{align*}
\Psi_{\alpha}  &  =\widehat{\mathcal{A}}\left\{  \Phi_{1}\left(  A_{1}\right)
\Phi_{2}\left(  A_{2}\right)  \Phi_{3}\left(  A_{3}\right)  f_{\alpha}\left(
\mathbf{x}_{\alpha},\mathbf{q}_{\alpha}\right)  \right\} \\
&  \simeq\widehat{\mathcal{A}}_{\beta\gamma}\left\{  \Phi_{\beta}\left(
A_{\beta}\right)  \Phi_{\gamma}\left(  A_{\gamma}\right)  ~g_{\alpha}\left(
\mathbf{x}_{\alpha}\right)  \right\}  \Phi_{\alpha}\left(  A_{\alpha}\right)
f_{\alpha}^{\left(  a\right)  }\left(  \mathbf{q}_{\alpha}\right) \\
&  =\Psi_{\alpha}^{\left(  2\right)  }\cdot\Phi_{\alpha}\left(  A_{\alpha
}\right)  f_{\alpha}^{\left(  a\right)  }\left(  \mathbf{q}_{\alpha}\right) \\
&  =\sum_{n_{y},l,n_{x},\lambda}C_{n_{y},l,n_{x},\lambda}^{\left(
\alpha\right)  }\left\vert n_{y},l,n_{x},\lambda;LM\right\rangle _{\alpha}\\
&  =\sum_{n_{x},\lambda}\sum_{n_{y},l}C_{n_{x},\lambda}^{\left(
\alpha\right)  }\cdot C_{n_{y},l}^{\left(  \alpha\right)  }\left\vert
n_{y},l,n_{x},\lambda;LM\right\rangle _{\alpha}.
\end{align*}
The two-cluster wave function $\Psi_{\alpha}^{\left(  2\right)  }$, and its
counterpart in oscillator space $\left\{  C_{n_{x},\lambda}^{\left(
\alpha\right)  }\right\}  $, is an eigenfunction of the two-cluster
Hamiltonian $\widehat{H}_{\alpha}^{\left(  2\right)  }$
\begin{equation}
\widehat{H}_{\alpha}^{\left(  2\right)  }=\sum_{i\in A_{\beta}+A_{\gamma}%
}\widehat{T}_{i}+\sum_{i<j\in A_{\beta}+A_{\gamma}}\widehat{V}\left(
ij\right)  . \label{eq:a016}%
\end{equation}
By solving the Schr\"{o}dinger equation
\begin{equation}
\sum_{\widetilde{n}_{x}=0}^{N_{2}}\left\langle n_{x},\lambda\left\vert
\widehat{H}_{\alpha}^{\left(  2\right)  }-E^{\left(  \alpha\right)
}\right\vert \widetilde{n}_{x},\lambda\right\rangle C_{\widetilde{n}%
_{x},\lambda}^{\left(  \alpha\right)  }=0 \label{eq:SchroMatr}%
\end{equation}
with a chosen number of \ basis functions $N_{2}$, we obtain the bound
state(s) $E_{\sigma}^{\left(  \alpha\right)  }$ ($\sigma=0,1,\ldots$) of the
two-cluster subsystem, which determine the threshold energy of the two-body
break up of the tree-cluster system.

In the second regime of the asymptotics we can neglect the antisymmetrization
operator and the short-range components of the inter-cluster potential. In
this regime, we use the Hyperspherical Harmonics (HH) basis to describe the
full decay of the three-cluster system, because (see, for instance,
\cite{kn:DanilZhuk93E, 1998PhRvC..58.1403C, kn:vasil97e, kn:ITP+RUCA2}) this
basis is the obvious choice for such type of three cluster behavior. The
transition from the bioscillator basis $\left\vert n_{y},l,n_{x}%
,\lambda;LM\right\rangle _{\alpha}$ to the HH basis $\left\vert n_{\rho
},K;l,\lambda;LM\right\rangle _{\alpha}$ (see details of the definition of HH
functions in e.g. \cite{kn:ITP+RUCA1}) is performed by an orthogonal matrix.
This transformation can be calculated in a straightforward way.

The asymptotic part of the wave function will then be represented by two sets
of expansion coefficients
\begin{equation}
\left\{  C_{n_{y},L}^{\left(  \alpha,E_{0}^{\left(  \alpha\right)  }\right)
}; \ C_{n_{\rho},K_{\min};L}^{\left(  \alpha\right)  },C_{n_{\rho},K_{\min
}+2;L}^{\left(  \alpha\right)  },\ldots,C_{n_{\rho},K_{\max};L}^{\left(
\alpha\right)  }\right\}  , \label{eq:a008}%
\end{equation}
where $K_{\min}=L$. All expansion coefficients in the asymptotic region have a
similar form, and consist of incoming ($\psi_{L}^{\left(  + \right)  }$,
$\psi_{K}^{\left(  + \right)  }$) and outgoing ($\psi_{L}^{\left(  -\right)  }
$, $\psi_{K}^{\left(  -\right)  }$) waves (see detail of the definition in
\cite{kn:ITP+RUCA2, kn:ITP+RUCA3}):%
\begin{align}
C_{n_{y},L}^{\left(  \alpha,E_{0}^{\left(  \alpha\right)  }\right)  }  &
\simeq\sqrt{2R_{n_{y}}}\left[  \delta_{c_{0}; \alpha}~\psi_{L}^{\left(
-\right)  }\left(  k_{\alpha}R_{n_{y}}\right)  -S_{c_{0}; \alpha}~\psi
_{L}^{\left(  +\right)  }\left(  k_{\alpha}R_{n_{y}}\right)  \right]
,\label{eq:a009}\\
C_{n_{\rho},K;L}^{\left(  \alpha\right)  }  &  \simeq\left(  2\rho_{n}\right)
^{2}\left[  \delta_{c_{0};\alpha K}~\psi_{K}^{\left(  -\right)  }\left(
k\rho_{n}\right)  -S_{c_{0};\alpha K}~\psi_{K}^{\left(  +\right)  }\left(
k\rho_{n}\right)  \right]  , \label{eq:a010}%
\end{align}
where the index $c_{0}$ denotes the entrance channel and%
\begin{align}
k  &  =\sqrt{\frac{2m}{\hbar^{2}}E},\quad\rho_{n}=b\sqrt{4n_{\rho}%
+2K+6}\label{eq:a011a}\\
k_{\alpha}  &  =\sqrt{\frac{2m}{\hbar^{2}}\left(  E-E_{0}^{\left(
\alpha\right)  }\right)  },\quad R_{n_{y}}=b\sqrt{4n_{y}+2L+3}.
\label{eq:a011b}%
\end{align}
Note that the index $\alpha$ numerates the binary channels, while both indexes
$\alpha$ and $K$ distinguish the ternary channels. Thus $c_{0}$ equals
$\alpha_{0}$, if the entrance channel is a binary one, or $c_{0}=\alpha
_{0},K_{0}$ for the three-cluster entrance channel.

With this definition of the asymptotic part of the wave function we deduce the
equations for the scattering parameters and wave function. By taking into
account (\ref{eq:SchroMatr}), (\ref{eq:a008}), (\ref{eq:a009}) and
(\ref{eq:a010}) one obtains%
\begin{align}
&  \sum_{\widetilde{\alpha}}\sum_{\widetilde{n}_{y},\widetilde{n}_{x}\leq
N_{i}}~_{\alpha}\left\langle n_{y},n_{x}\left\vert \widehat{H}-E\right\vert
\widetilde{n}_{y},\widetilde{n}_{x}\right\rangle _{\widetilde{\alpha}%
}C_{\widetilde{n}_{y},\widetilde{n}_{x}}^{\left(  \widetilde{\alpha}\right)
}\nonumber\\
&  -\sum_{\widetilde{\alpha}}S_{c_{0};\widetilde{\alpha}}~\sum_{\widetilde
{n}_{y}>N_{i}}~_{\alpha}\left\langle n_{y},n_{x}\left\vert \widehat
{H}-E\right\vert \widetilde{n}_{y},E_{0}^{\left(  \widetilde{\alpha}\right)
}\right\rangle _{\widetilde{\alpha}}~\psi_{L}^{\left(  +\right)  }\left(
k_{\widetilde{\alpha}}R_{\widetilde{n}_{y}}\right) \nonumber\\
&  -\sum_{\widetilde{\alpha}}S_{c_{0};\widetilde{\alpha}\widetilde{K}}%
~\sum_{\widetilde{n}_{\rho}>N_{i}}~_{\alpha}\left\langle n_{y},n_{x}\left\vert
\widehat{H}-E\right\vert \widetilde{n}_{\rho},\widetilde{K}\right\rangle
_{\widetilde{\alpha}}~\psi_{\widetilde{K}}^{\left(  +\right)  }\left(
k\widetilde{\rho}\right)  =\label{eq:a012}\\
&  -\sum_{\widetilde{\alpha}}\delta_{c_{0};\widetilde{\alpha}}~\sum
_{\widetilde{n}_{y}>N_{i}}~_{\alpha}\left\langle n_{y},n_{x}\left\vert
\widehat{H}-E\right\vert \widetilde{n}_{y},E_{0}^{\left(  \widetilde{\alpha
}\right)  }\right\rangle _{\widetilde{\alpha}}~\psi_{L}^{\left(  -\right)
}\left(  k_{\widetilde{\alpha}}R_{\widetilde{n}_{y}}\right) \nonumber\\
&  -\sum_{\widetilde{\alpha}}\delta_{c_{0};\widetilde{\alpha}\widetilde{K}%
}~\sum_{\widetilde{n}_{\rho}>N_{i}}~_{\alpha}\left\langle n_{y},n_{x}%
\left\vert \widehat{H}-E\right\vert \widetilde{n}_{\rho},\widetilde
{K}\right\rangle _{\widetilde{\alpha}}~\psi_{\widetilde{K}}^{\left(  -\right)
}\left(  k\widetilde{\rho}\right)  .\nonumber
\end{align}
This system of linear equations contains in\ the general case of three
different clusters
\begin{equation}
\frac{3}{2}\left(  N_{i}+1\right)  \left(  N_{i}+2\right)  +3+3N_{ch.HH}%
\end{equation}
parameters to be determined. Here%
\begin{equation}
N_{ch.HH}=\left(  K_{\max}-K_{\min}\right)  /2+1
\end{equation}
is the number of HH channels. From this total amount, $\frac{3}{2}\left(
N_{i}+1\right)  \left(  N_{i}+2\right)  $ coefficients represent the wave
function in the internal region, and the other $3+3N_{ch.HH}$ parameters
determine the elastic and inelastic processes, leading to two or three
clusters in the exit channels. These parameters unambiguously define the wave
functions in the asymptotic region.

\section{Results}

We use the Minnesota (MP) \cite{kn:Minn_pot}, and the modified Hasegawa-Nagata
(MHNP) \cite{potMHN1, potMHN2} nucleon-nucleon ($NN$) potentials. The
oscillator radius $b$ is chosen to optimize the bound state energy of the
deuteron, and equals $b=1.512$ fm for MP and $b=1.668$ fm for MHNP.
Considering these two potentials reveals the effect of peculiarities of
$NN$-forces on the parameters of resonance states.

In a first calculation, we neglect all binary channels, and only consider the
three-cluster channels. This allows to understand what kind of resonances are
generated by this channel only. We have omitted spin-orbital components of the
$NN$-forces to reduce the computational burden.
Results obtained in this approximation can be considered to represent a
\textquotedblleft lower limit\textquotedblright\ for the width of a resonance,
as additional channels will open new decay possibilities of the resonance,
which will increase its width. In this respect the three-cluster approximation
will indicate whether some resonances state(s) could survive after binary
channels are included.

By solving the dynamic equations (\ref{eq:a012}) for $N_{c}$ channels, we
directly obtain the $N_{c}\times N_{c}$ $S$-matrix. There are two different
parametrizations for the $S$-matrix. In the first one, each element $S_{ij}$
can be represented by the phase shift $\delta_{ij}$ and the inelastic
parameter $\eta_{ij}$: $S_{ij}=\eta_{ij}\exp\left\{  2i\delta_{ij}\right\}  $.
In the second parametrization the $S$-matrix is reduced to diagonal form by an
orthogonal transformation:%
\begin{equation}
\left\Vert S\right\Vert =\left\Vert O\right\Vert ^{T}\cdot\left\Vert
S^{\left(  E\right)  }\right\Vert \cdot\left\Vert O\right\Vert .
\end{equation}
The latter procedure leads to $N_{c}$ uncoupled elastic \textquotedblleft
eigenchannels\textquotedblright, whose (eigen)phase shifts parametrize the
diagonalized $S^{\left(  E\right)  }$-matrix. We use the eigenphase shifts to
determine both the energy and width of the resonances. They are obtained from
the following equations%
\begin{equation}
\frac{d^{2}\delta_{\nu}}{d^{2}E}|_{E=E_{r}}=0,\quad\Gamma=2\left(  \frac
{dE}{d\delta_{\nu}}\right)  |_{E=E_{r}}.
\end{equation}

\begin{figure}
[ptb]
\begin{center}
\includegraphics[
height=11.5257cm,
width=16.277cm
]%
{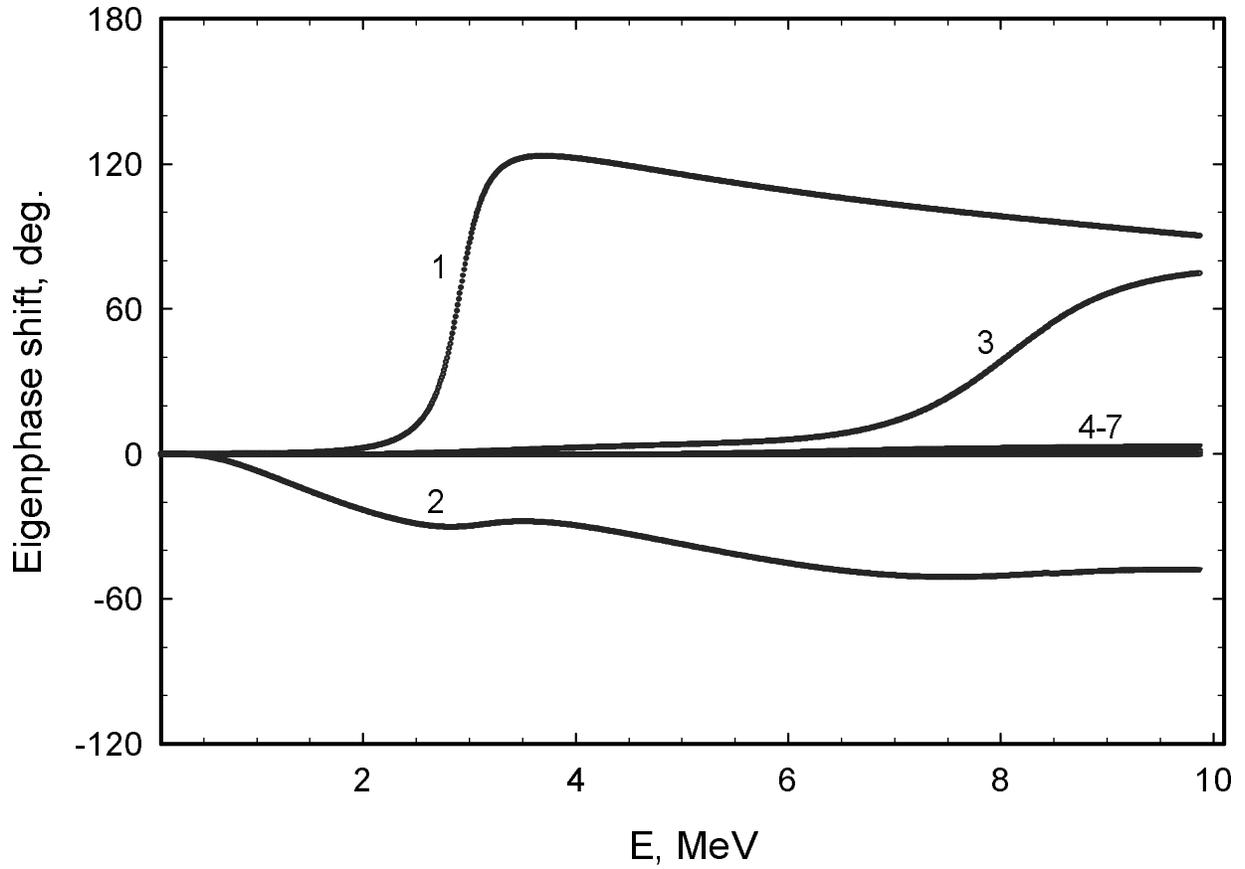}%
\caption{Eigenphase shifts for the $L^{\pi}=1^{-}$, $S=1$ state of $^{4}Li$
obtained with the Minnesota potential and $K_{max}=11$. The eigenchannels are
numerated.}%
\label{Fig:4Li_eigenphases_L1_S1}%
\end{center}
\end{figure}

We start our analysis from eigenphase shift. In Fig.
\ref{Fig:4Li_eigenphases_L1_S1} we display eigenphase shift
of so-called 3$\Rightarrow$3 scattering for $L^{\pi}=1^{-}$, $S=1$ state in
the $^{4}Li$, obtained with the Minnesota potential. Similar pictures are
obtained for other nuclei and different ($L^{\pi}$, $S$) states and also with
the MHN potential. One can see from Fig. \ref{Fig:4Li_eigenphases_L1_S1} that
there are two resonance states in $^{4}Li$, first resonance is narrow and
manifest itself through the first eigenchannel, while the second resonance is
very broad and appear in the third eigenchannel.

The eigenphase shifts provide with a direct and simple way to determine the
energy and width of a resonance state in the three-cluster continuum, but to
get information concerning the main features of the three-cluster dynamics one
needs to analyze the phase shifts $\delta_{ij}$ and the inelastic parameters
$\eta_{ij}$.

\begin{figure}
[ptb]
\begin{center}
\includegraphics[
height=11.1413cm,
width=16.2111cm
]%
{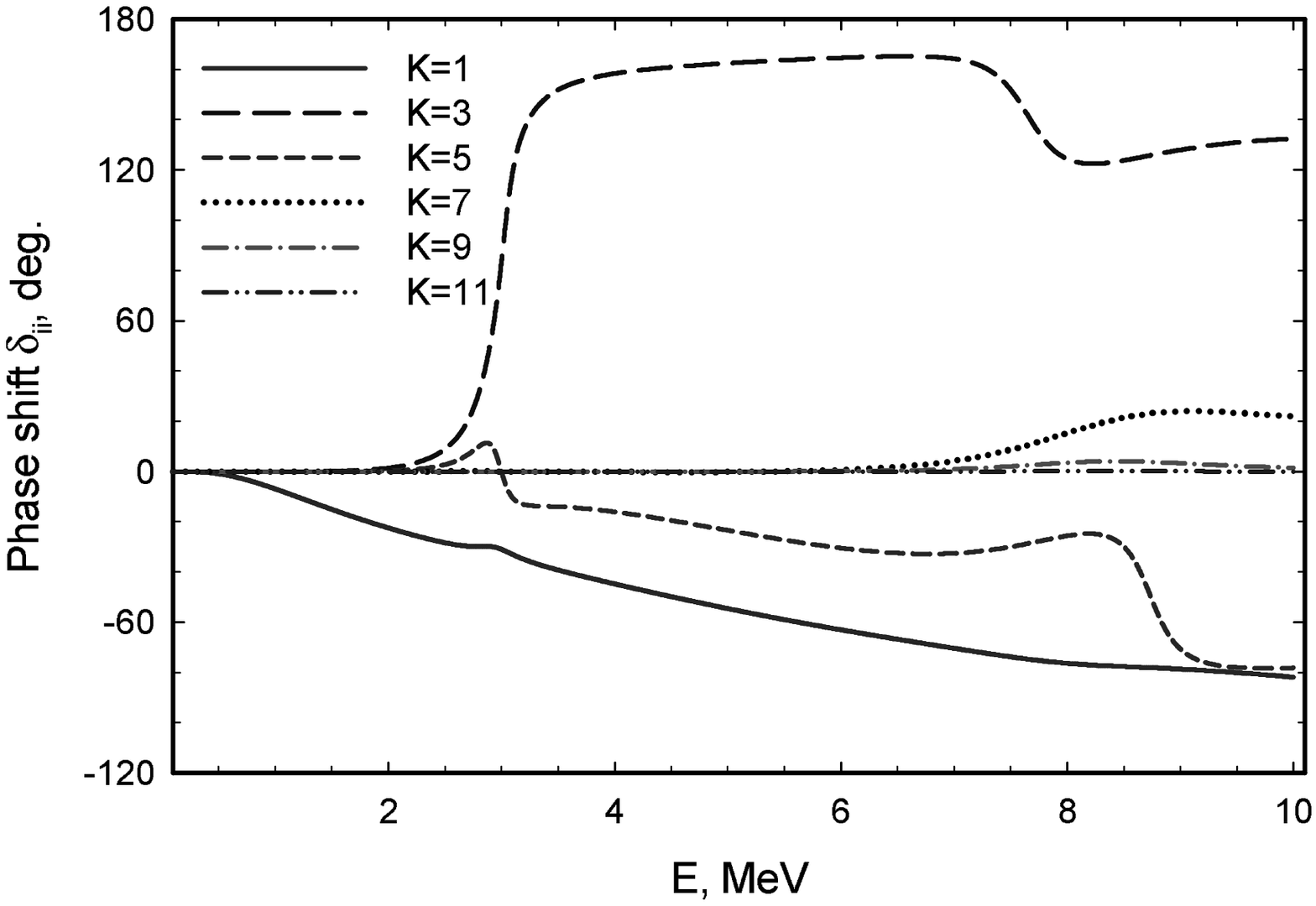}%
\caption{Phase shifts of the 3$\Rightarrow$3 scattering for the $L^{\pi}%
=1^{-}$, $S=1$ state of $^{4}Li$. Results are obtained with the Minnesota
potential and $K_{max}=11$.}%
\label{Fig:4Li_phases_L1_S1}%
\end{center}
\end{figure}

\begin{figure}
[ptbptb]
\begin{center}
\includegraphics[
height=11.0688cm,
width=16.02cm
]%
{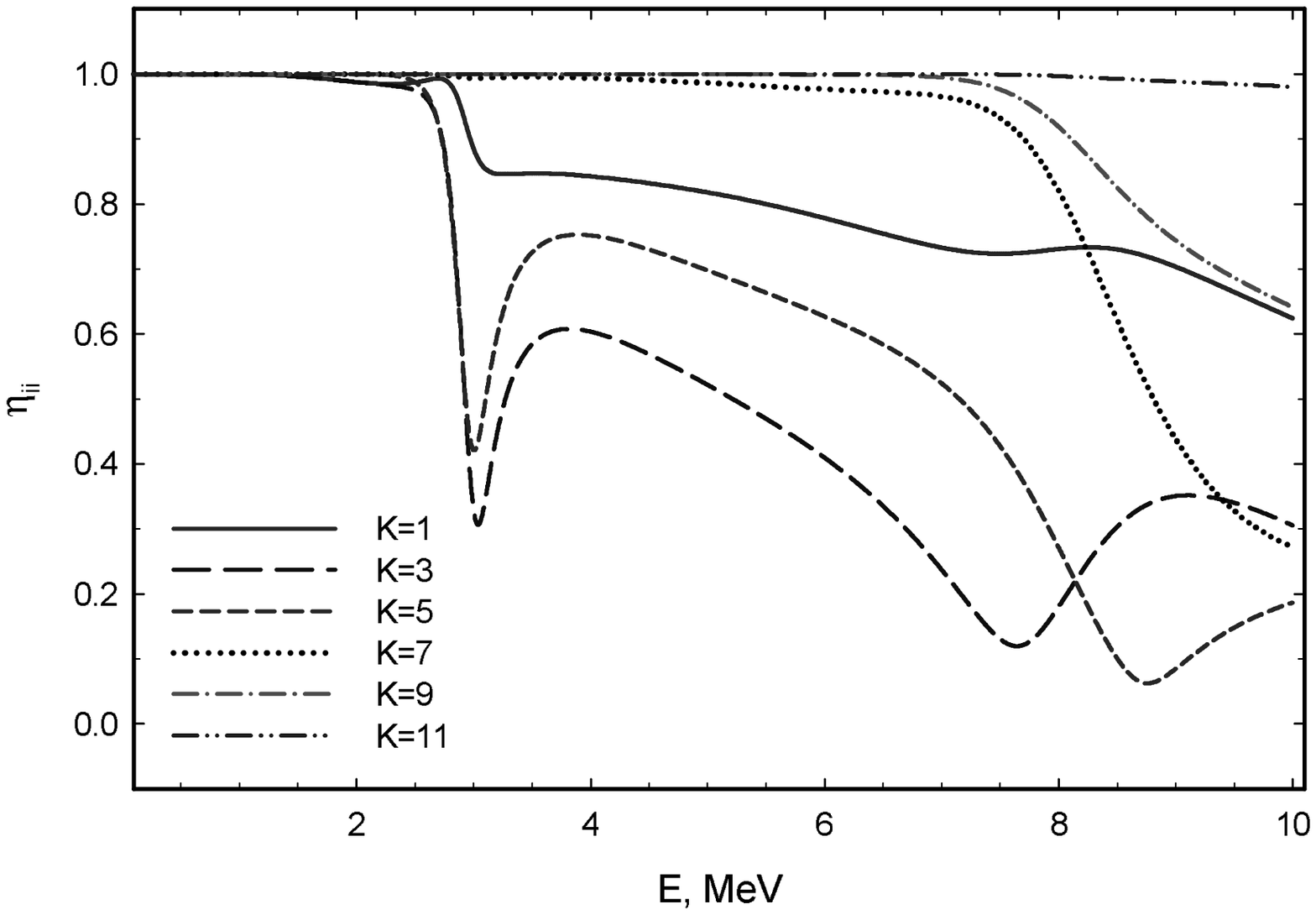}%
\caption{Inelastic parameters of the 3$\Rightarrow$3 scattering for the
$L^{\pi}=1^{-}$, $S=1$ state in $^{4}Li$. Results are obtained with the
Minnesota potential and $K_{max}=11$.}%
\label{Fig:4Li_etas_L1_S1}%
\end{center}
\end{figure}

In Figs. \ref{Fig:4Li_phases_L1_S1} and \ref{Fig:4Li_etas_L1_S1} we display
the phase
shifts $\delta_{ii}$ and inelastic parameters $\eta_{ii}$
connected with the diagonal matrix elements of the original $S$-matrix. They
are obtained for $^{4}Li$ and with MP. One notices that only two
hyperspherical harmonics are responsible for the lowest $1^{-}$ resonance
state. The $K=3$ phase shift displays the classical behavior for a resonance
state, while the $K=5$ phase shift indicates a \textquotedblleft
shadow\textquotedblright\ resonance. Many more hyperspherical momenta are
involved in creating the second resonance.

In Tables \ref{Tab:ThreeClusterResonances MP} and
\ref{Tab:ThreeClusterResonances MHNP} we collect the parameters of\ the
resonance states lying above the three-cluster threshold $d+N+N$. The even
parity states are obtained with $K_{\max}=10$, and the odd parity states with
$K_{\max}=11$.%

\begin{table}[h] \centering
\begin{tabular}
[c]{|c|c|c|c|c|c|c|c|}\hline
Nucleus & $L^{\pi}$ & $S$ & $K_{\max}$ & $E$, MeV & $\Gamma$, MeV & $E$, MeV &
$\Gamma$, MeV\\\hline
$^{4}H$ & $1^{-}$ & 0 & 11 & 1.642 & 0.367 & 6.726 & 2.759\\\hline
$^{4}H$ & $1^{-}$ & 1 & 11 & 1.911 & 0.374 & 6.958 & 2.982\\\hline
$^{4}Li$ & $1^{-}$ & 0 & 11 & 2.604 & 0.413 & 7.787 & 3.141\\\hline
$^{4}Li$ & $1^{-}$ & 1 & 11 & 2.912 & 0.465 & 8.085 & 3.384\\\hline
$^{4}He$ & $2^{+}$ & 0 & 10 & 1.950 & 0.233 & 2.904 & 0.207\\\hline
\end{tabular}
\caption{ Resonance states of $^{4}H$,  $^{4}He$ and $^4Li$,
created by the three-cluster channel $d+N+N$. Results are obtained
with the Minnesota potential. }\label{Tab:ThreeClusterResonances MP}
\end{table}
\begin{table}[h] \centering
\begin{tabular}
[c]{|c|c|c|c|c|c|c|c|}\hline
Nucleus & $L^{\pi}$ & $S$ & $K_{\max}$ & $E$, MeV & $\Gamma$, MeV & $E$, MeV &
$\Gamma$, MeV\\\hline
$^{4}H$ & $1^{-}$ & 0 & 11 & 3.972 & 1.170 & 9.469 & 3.440\\\hline
$^{4}H$ & $1^{-}$ & 1 & 11 & 3.738 & 0.950 & 9.250 & 3.362\\\hline
$^{4}Li$ & $1^{-}$ & 0 & 11 & 0.748 & 0.093 & 5.009 & 1.531\\\hline
$^{4}Li$ & $1^{-}$ & 1 & 11 & 0.662 & 0.056 & 4.772 & 1.329\\\hline
$^{4}He$ & $2^{+}$ & 0 & 10 & 0.890 & 0.005 & 2.436 & 0.167\\\hline
\end{tabular}
\caption{ Resonance states of $^{4}H$,  $^{4}He$ and $^4Li$,
created by the three-cluster channel $d+N+N$. Results are obtained
with the MHN potential. }\label{Tab:ThreeClusterResonances MHNP}
\end{table}%
%

It is known that there is an effective barrier in each channel of
three-cluster system. The barrier is created by a potential well, resulted
from a NN-interaction between nucleons from different clusters, and
centrifugal barrier, which is proportional to
\begin{equation}
\frac{\hbar^{2}}{2m}\frac{K(K+4)}{\rho^{2}}\nonumber
\end{equation}
In $^{4}He$ and $^{4}Li$ the Coulomb repulsion
\begin{equation}
\frac{Z_{eff}}{\rho}\nonumber
\end{equation}
have to be added to the effective barrier. The effective charge $Z_{eff}$
depends on quantum numbers $\alpha$, $K$, $l_{1}$, $l_{2}$ and its definition
can be found in \cite{kn:ITP+RUCA1}. The deeper is potential well, the larger
is the effective barrier and, consequently, more resonance states can be
created by such effective barrier. One can see that the effective barrier,
generated by the MHN potential, is more deeper than the one connected with the
Minnesota potential. As a result of this difference, the resonance states in
$^{4}H$, obtained with the MHN potential, have larger energy then resonances,
calculated with the Minnesota potential.

\begin{figure}
[ptb]
\begin{center}
\includegraphics[
height=4.3094in,
width=5.7709in
]%
{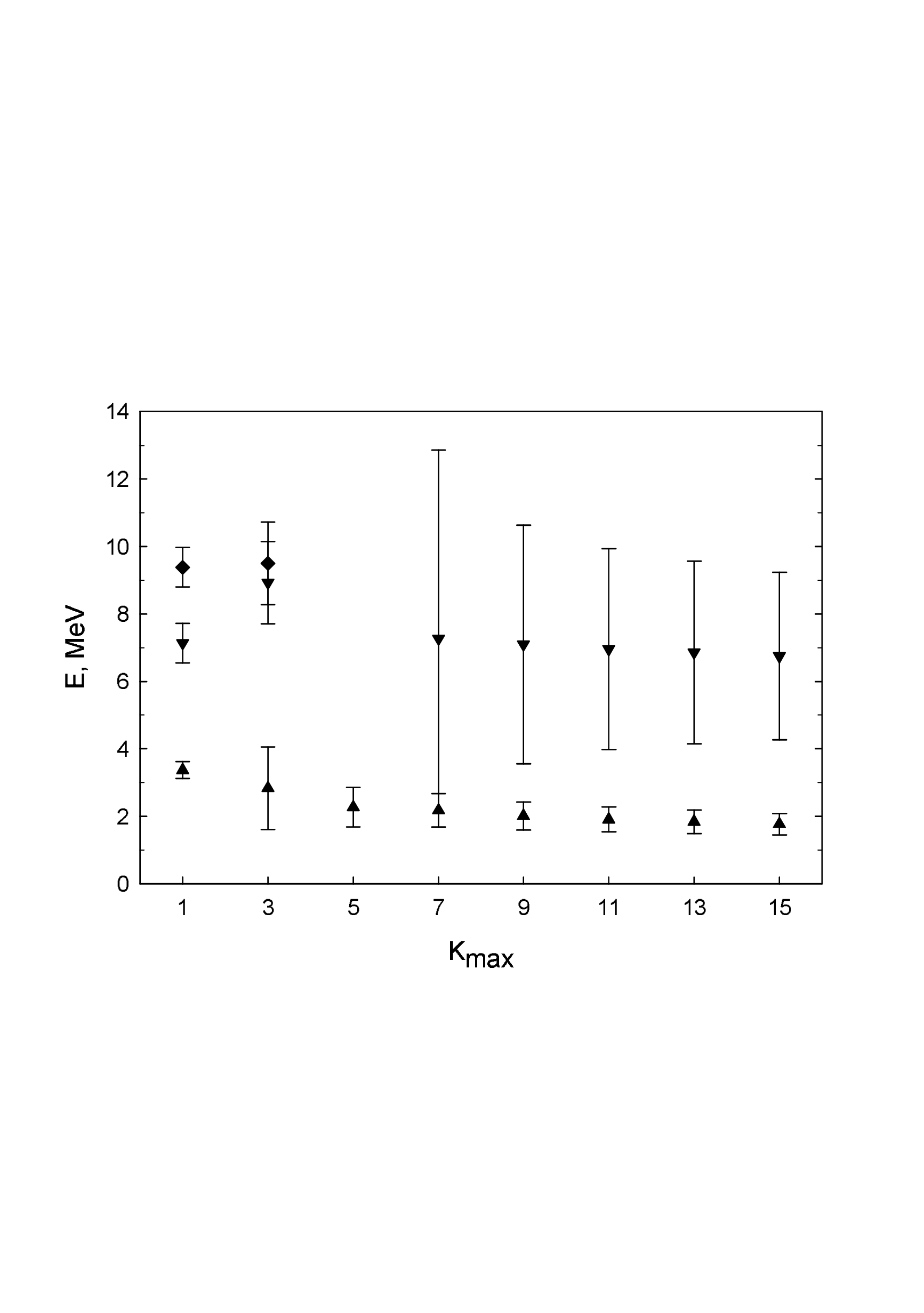}%
\caption{Energy of the $1^{-}$ resonance state of $^{4}H$ (total spin $S=1$)
as a function of $K_{max}$. Error bars indicate the double of the resonance
width. The calculations have been performed with the Minnesota potential.}%
\label{Fig:H4_resonances_vs_Kmax}%
\end{center}
\end{figure}

As we pointed out the results, presented in Tables
\ref{Tab:ThreeClusterResonances MP} and \ref{Tab:ThreeClusterResonances MHNP},
for the even parity states are obtained with $K_{\max}=10$, and for the odd
parity states with $K_{\max}=11$. These values for $K_{\max}$ are\ sufficient
to obtain stable results. This is demonstrated in Fig.
\ref{Fig:H4_resonances_vs_Kmax}
where the parameters (energy and width) of the $1^{-}$ resonance in $^{4}H$
are displayed as a function of $K_{\max}$. The results in Fig.
\ref{Fig:H4_resonances_vs_Kmax} are presented for the Minnesota potential, and
similar results are obtained for the MHN potential. We indicate some
\textquotedblleft false\textquotedblright\ resonances states appearing at
small values of $K_{\max}$ due to the restriction on decay channels compatible
with this $K_{\max}$. When we increase the number of open channels, the
\textquotedblleft false\textquotedblright\ resonances disappear and the
physical resonances converge to their correct positions.

There are some arguments that the physical resonances do not depends on the
boundary conditions implemented, or on the used approximations. One can e.g.
study the behavior of the so-called Harris states, i.e. the eigenvalues of the
Hamiltonian as a function of the number of basis functions involved in a
calculation. It was shown (see, for instance, \cite{kn:cohstate2E, kn:CSM-Ho})
that the eigenvalues $E_{\nu}\left(  n\right)  $ ($E_{\nu}\left(  n\right)  $
is $\nu$-th eigenvalue, obtained with $n$ basis functions) create plateaus at
the energies of resonance states. For a very narrow resonance this plateau is
already observed for a small value of $n$. For wider resonances, one needs
more basis functions to reach a plateau. For a small number of basis functions
a wider resonance can become apparent as the repulsion of two eigenvalues
(avoided crossing of two eigenvalues). Such behavior of these eigenvalues was
observed for the Hamiltonian of the three-cluster configuration $d+N+N$ in
$^{4}H$, $^{4}He$ and $^{4}Li$.

\section{Conclusion}

A microscopic model is formulated to treat properly the two- and three-body
boundary conditions. For this aim the Faddeev component is used. The
hyperspherical harmonics are used to numerate three-cluster channels. They are
very valuable for describing three-cluster asymptotic. Two \textit{NN}%
-potentials are involved in the calculations in order to evaluate the
sensitivity of the final results with respect to this important factor of a
microscopic model.

The model is applied for studying resonances states in $^{4}H$, $^{4}He$ and
$^{4}Li$ nuclei, created by the three-cluster configuration $d+N+N$. The
results presented here demonstrate that the three cluster configuration
creates an effective centrifugal barrier which allows to accommodate several
resonances. The effect of the two-cluster channels on the position and width
of the three-cluster resonances will be discussed in future work.

\section{Acknowledgments}

One of the authors (V. V.) gratefully acknowledges the University of Antwerp
(RUCA) for a \textquotedblleft RAFO-gastprofessoraat
2002-2003\textquotedblright\ and the kind hospitality of the members of the
research group \textquotedblleft Computational Modeling and
Programing\textquotedblright\ of the Department of \textquotedblleft
Mathematics and Computer Sciences\textquotedblright, University of Antwerp,
RUCA, Belgium.

\end{document}